# Communication Bias in Large Language Models: A Regulatory Perspective

ADRIAN KUENZLER, is an Associate Professor at the Faculty of Law, University of Hong Kong, Hong Kong

STEFAN SCHMID, is a Professor at the Faculty of Electrical Engineering and Computer Science, TU Berlin, Germany, and a Principal Investigator at the Weizenbaum Institute in Berlin, Germany[1]

Large language models (LLMs) have become integral to numerous applications, raising concerns about bias, fairness, and compliance with emerging regulatory frameworks. This paper provides a review of some of the most significant risks associated with biased LLM outputs and their broader societal implications. We discuss how regulatory initiatives such as the European Union's (EU) Artificial Intelligence Act (AI Act) and the Digital Services Act (DSA) seek to address these challenges by introducing legal requirements for transparency, accountability, and harm mitigation. In addition, we explore various approaches for assessing and mitigating bias, along with the limitations of current methods. While regulatory frameworks play a crucial role in governing artificial intelligence (AI), they also present challenges in balancing innovation with legal compliance. As oversight mechanisms continue to evolve, we emphasize the fact that LLMs are increasingly regarded as a distinct subset of AI systems and that LLMs are particularly noteworthy due to their immediate interaction with consumers, functioning as the primary interface between users and AI. This distinguishes LLMs from many other AI applications operating behind the scenes and renders them unique in the sense that they are becoming integral tools for personal decision-making, offering validation, guidance, and support across various aspects of everyday life, including healthcare, finance, and even choices made in relation to politics. While the current landscape of AI-generated content is addressed by several new regulatory frameworks, none of them have been devised with the possibility in mind that LLMs may affect citizens' fundamental worldviews and social perspectives, including their individual and collective voting decisions. Our analysis highlights the need for a greater focus on both competition and technology design governance as a complement to value chain and content-based regulation, in order to foster fair and trustworthy AI systems.

**KEY INSIGHTS**

- **LLMs are central to modern communication**—while they enhance productivity and improve decision-making, their character as the primary interface between users and AI enables them to subtly influence individual and collective opinions, including democratic processes.
- **As AI systems become gatekeepers of information**, LLMs may reinforce existing biases and promote particular perspectives, creating and perpetuating echo chambers and intensifying societal polarization.
- **Emerging regulatory frameworks such as the EU AI Act and the DSA** seek to mandate transparency, accountability, and fairness in AI systems, yet they often address communication bias only as a byproduct of broader safety and content moderation measures.
- **Detecting and mitigating communication bias in LLMs is inherently challenging** due to the models' complexity and the subtle, multifaceted nature of such bias.
- **A comprehensive approach combining value chain regulation, content moderation, competition, and ongoing technology design governance** is crucial in fostering diverse and transparent AI systems that mitigate bias while promoting a balanced digital information ecosystem.

## 1 INTRODUCTION

Artificial intelligence (AI) involves computational systems designed to perform tasks that traditionally require human intelligence, such as perception, reasoning, and language understanding. Large language models (LLMs) are a prominent subset of AI, built on advanced neural network architectures that can generate new data, including text, images, and audio. LLMs utilize various technologies to identify patterns in a given set of training data, without requiring explicit instructions about what to look for [12, 35]. LLMs typically assume that the training data follows a probability distribution, and once they have identified existing patterns, they can generate new instances that are similar to the original data. By drawing from and combining training data, LLMs can create new content that transcends the initial dataset [17]. This process leads to technical limitations: for instance, LLMs can hallucinate by producing plausible yet factually incorrect or fabricated information; LLMs can struggle with ambiguous or missing data, leading to blind spots in their outputs; and LLMs can reflect and amplify existing preconceptions present in their algorithms or training data, leading to communication bias.

We propose a broad definition of communication bias—the expression, amplification, and systematic favoring of certain social, cultural, or political perspectives in the output generated by LLMs, which can affect the attitudes and beliefs of users and thus, ultimately, public discourse. We use the term communication bias as opposed to data bias, a possible cause of communication bias, arising from skewed or unrepresentative training data, and automation bias, which may exacerbate these effects by making users more likely to accept and act on biased AI outputs without sufficient scrutiny. Specifically, communication bias need not only manifest in the selective presentation of facts or the rendering of false or made-up information, but can also reside in claims that sit somewhere in between, or outside, the categories of being either false or truthful [22]. Here, a prejudiced effect results from generative expression rather than imbalances in training data (data bias) or uncritical user uptake (automation bias). Recent research shows that commercial LLMs consistently display communication bias and can adapt their responses to user or persona prompts [3]. The integration of LLMs into consumer-facing applications such as chatbots, virtual assistants, or search engines has rendered them the primary interfaces between users and AI, enabling immediate, interactive communication and content generation. Hence, LLMs have a broad range of applications across various sectors [9]. They can be employed to assist businesses by analyzing financial data and generating content, and can be used for tasks such as text generation, translation, and summarization. LLMs can also enhance personal user experiences. For instance, LLMs can facilitate patient–doctor communications; aid in personalized learning; provide life advice; and inform political decisions [37]. Owing to these features, LLMs have the potential to supplant existing gatekeepers and to become agents that mediate various interactions, select underlying applications, and choose providers to respond to user queries [1]. By assuming control and entrusting consumers' ability to steer demand, input, and data to AI agents, LLMs may gradually come to serve as the main access points by which users will engage with digital technologies and the Internet, eluding the grasp of present legal frameworks and thus turning them into a distinct and vital new object of regulatory consideration [23].

This paper gives an overview of the rapidly developing legal and regulatory background that governs AI and machine learning technologies, with a particular emphasis on the ability of LLMs to affect citizens' political opinions and voting decisions. Recent studies demonstrate that LLMs can influence people's views, but there is currently no framework that can effectively address the issue. As AI and machine learning technologies become more widespread, there are increasingly urgent calls for proactive regulatory frameworks to establish guardrails that address the models' limitations and ensure their responsible use within society [13]. For instance, the FBI's 2023 Internet Crime Report highlights that manipulation and social engineering have led to substantial financial losses, surpassing USD 12.5 billion in 2023 [11]. The European Union Agency for Cybersecurity's 2023 Threat Landscape Report identifies highly realistic synthetic media as progressively more hazardous to digital identity and trust [10]. Due to the inherent complexity of LLMs and the difficulty of targeting communication bias in the models' outputs, the search for new regulatory approaches remains critical in ensuring an open digital information ecosystem [3].

An emerging source of tension resides in the possibility that LLMs carry communication biases. For example, studies have revealed a variety of political leanings across different models and have demonstrated how such biases can impact the effectiveness of hate speech and misinformation detection systems [3]. Other studies have identified biases in open-source LLMs regarding divisive topics such as immigration, reproductive rights, and climate change [3]. While most existing models appear to have liberal dispositions and are primarily US-centered in their outputs, LLMs can also exhibit a range of conservative biases, depending on the issue [3]. Such biases can be attributed to the models' algorithms and/or training data, but they may also arise as a result of purposeful decisions by AI companies. Users may not notice any bias when it comes to the deployment of LLMs, and most models have built-in safeguards designed to prevent the generation of toxic content. However, communication bias can emerge in subtle ways, and the additional training aimed at limiting certain kinds of content generation can inadvertently introduce more bias. The problem may escalate as LLMs become more widespread, with future generations of the models increasingly trained on data that includes AI-generated output [14]. In addition, some social groups may try to sway LLMs to promote their perspectives over those of others and may attempt to influence the models' training data. The issue is exacerbated if a few influential corporations come to dominate the AI market, providing citizens with the information that creates public knowledge [7]. If LLMs selectively present or omit particular perspectives, this will inevitably shape public discourse by limiting the diversity of available opinions and possibly undermining democratic processes.

A number of regulations in the EU seek to address this tension. However, none of them applies across the board, and the targeting of subtle communication bias in the output generated by LLMs frequently turns out to be a byproduct of more straightforward mechanisms intended to ban illegal content. In the following sections, we give an overview of the existing regulations and offer potential routes for further discussion about the risks of communication bias in LLMs and how such risks might best be addressed. While our focus is on the AI Act and the DSA, these frameworks do not operate in isolation: the size of the EU market and the EU's comprehensive and enforceable legal standards set de facto global norms that companies around the world must increasingly follow. AI technologies, especially LLMs, are developed and deployed transnationally, and different legal frameworks may lead to regulatory fragmentation, compliance burdens, and uneven protection for users [13]. At this stage, focusing predominantly on the EU is justified because its rules are already influencing legislative agendas and business practices in other jurisdictions, serving as a blueprint for global AI governance. Examining the EU's regulatory model and its underlying principles provides unique insight into the evolving global environment, the challenges of harmonization, and the practical realities faced by AI providers.

## 2 AI-ENABLED COMMUNICATION: RISKS AND MITIGATION STRATEGIES

Several studies have focused on identifying bias present in the output of LLMs, which can lead to subtle but harmful stereotypes and misinformation [3]. In this section, we elaborate on some specific challenges relating to such bias and how these challenges complicate the backdrop of trustworthiness and reliability of AI-enabled communication.

### 2.1 Risks of AI-Enabled Communication

With the increasing spread of AI-generated outputs, digital information ecosystems are changing in a number of ways. A relatively well-known and distinct issue is that providers of AI systems may generate synthetic content as well as deepfakes or AI-generated text that can mislead individuals and the public and erode trust in AI-driven communication (Art. 50 AI Act). A far less overt form of harm that may nonetheless corrode the conditions of free and equitable public discourse concerns the generation of a vast amount of content that aligns with certain political narratives or worldviews. In the digital age, this threat is characterized by the shift from state-controlled speech to platform-dominated governance, where algorithms often prioritize engagement over democratic integrity [31]. As LLMs become integrated into content creation processes or online services such as search engines or chatbots, their output may subtly influence popular attitudes and beliefs and may impact societal perceptions. AI-powered search engines offer more conversational and context-aware interactions compared to traditional ones and are better able to personalize and customize their outputs based on user interactions [23, 26]. When responses are produced that align with user preferences rather than objective information, this phenomenon can contribute to the creation and perpetuation of echo chambers and result in increasing polarization. In turn, AI-generated outputs can facilitate feedback loops that reinforce existing beliefs and attitudes and can limit exposure to a diversity of perspectives. Even if users are aware that they are relying on AI-generated content, they may still be prone to contributing to societal polarization and less informed decision-making because LLMs can imperceptibly direct public opinion toward particular party positions and produce an unbalanced representation of views on public affairs [26]. In this way, LLMs increasingly mediate the communicative acts that are deemed necessary for the formation of public opinion by curating information, yet their biases risk distorting public discourse, reinforcing echo chambers, marginalizing dissent, and undermining democratic authorship [31].

### 2.2 Current Approaches to Address the Risks of AI-Enabled Communication

Due to the subtle nature of AI-enabled communication, efforts to address its risks will likely rely on a multifaceted approach based on what we refer to as value chain regulation, content moderation, competition, and technology design governance. Legislators, particularly in the EU, have established ex ante requirements across the AI value chain under the AI Act, imposing legal obligations on several actors, depending on the role they play. The AI Act targets providers (those who develop LLMs or make them available), deployers (those who use LLMs), importers, and distributors. It introduces a risk-based approach, categorizing AI systems according to different risk levels—unacceptable, high-risk, limited risk, and minimal risk—with stricter obligations for high-risk and general-purpose AI (GPAI) models, including LLMs (Art. 51(1)(a), Annex XIII AI Act). Some key requirements include the establishment of risk management systems (Art. 9), data governance and quality criteria (Art. 10), the provision of technical documentation (Art. 11), transparency obligations (Art. 13), and human oversight (Art. 14). For GPAI and high-impact GPAI models, additional transparency, documentation, testing, evaluation and risk mitigation duties apply (Arts. 53 and 55 AI Act, Recitals 110-111 AI Act). Enforcement is carried out by national market surveillance authorities and the EU AI Office, with fines for noncompliance reaching up to 7% of global annual turnover (Art. 99 AI

Act). These measures seek to ensure that AI systems are developed and deployed with attention to fairness, transparency, and bias mitigation, especially where outputs may affect fundamental rights or democratic processes. By contrast, content moderation, particularly under the DSA, operates ex post, focusing on outputs. The DSA does not specifically address LLMs; it regulates online intermediaries, including platforms and search engines, targeting harmful or illegal content as well as advertising transparency (Art. 1 DSA). Even so, the DSA's provisions apply to LLMs when LLMs are integrated into platforms covered by the DSA. For instance, if LLMs are used for content generation or moderation on very large online platforms (VLOPs) or very large online search engines (VLOSEs), these platforms must ensure compliance with the DSA (Art. 33 DSA). In particular, such platforms must implement mechanisms for identifying and mitigating systemic risks, including those related to illegal content, civic discourse, and fundamental rights (Art. 34 DSA). Furthermore, the DSA mandates transparency reporting (Art. 15), user redress mechanisms (Art. 20), recommender system disclosures (Art. 27), and crisis response protocols (Art. 36). Enforcement is managed by Digital Services Coordinators and the European Commission (EC), with fines for violations up to 6% of global annual turnover (Art. 74 DSA). While the DSA's scope is narrower—concentrating on platforms rather than the entire AI value chain—it may nonetheless be critical for managing, through human and/or automated filtering systems, the societal impact of LLM-generated content.

From a conceptual perspective, value chain regulation (AI Act) and content moderation (DSA) differ in their objectives, scope of application, and methods of enforcement. Content moderation is typically more limited in scope than value chain regulation and is conducted through a combination of automated systems and human oversight. Both value chain regulation and content moderation are capable, however, of addressing bias mitigation, at least to a certain degree. Value chain regulation seeks to mitigate bias primarily through pre-market mechanisms, by ensuring high-quality data sets for training, bias testing, and accuracy requirements. The emphasis of content moderation is on post-market instruments such as the removal of harmful or illegal content, algorithmic transparency, and user complaints. While value chain regulation proactively governs system architecture and deployment, content moderation focuses reactively on managing outputs. Neiter, however, addresses how the development and use of LLMs are influenced by the market mechanism itself. We therefore emphasize the need for complementary strategies, including competition policy and ongoing technology design governance, to keep a broader focus on the incentives that shape AI-mediated communication and how these incentives work to foster a pluralistic and trustworthy digital information ecosystem.

| | Value chain regulation (AI Act) | Content moderation (DSA) |
|---|---|---|
| Primary objective | Establish ex ante requirements for AI development/deployment | Manage ex post outputs to prevent dissemination of harmful/illegal content |
| Regulatory scope | All actors in the AI value chain (providers, deployers, importers, distributors) | Intermediary services (VLOPs/VLOSEs hosting LLM-generated content) |
| Risk framework | Risk-based tiers: unacceptable > GPAI > high-risk > limited risk > minimal risk | Systemic risk mitigation for VLOPs/VLOSEs (illegal content, civic discourse, basic rights) |
| Key obligations | Conformity assessments<br>Certain fundamental rights impact assessments<br>Technical documentation<br>Human oversight | Notice-and-action mechanisms<br>Transparency reporting<br>Recommender system disclosures<br>Crisis response protocols |
| Enforcement | Fines up to 7% global annual turnover<br>National market surveillance authorities<br>EU-wide database for high-risk systems | Fines up to 6% global annual turnover<br>Digital Services Coordinators<br>Independent audits |
| Bias mitigation | Pre-market: training data governance, bias testing, accuracy requirements | Post-market: content removal, algorithmic transparency, user redress mechanisms |
| Differences | Ex ante compliance focused on system architecture | Ex post accountability focused on output management |

Table 1. Regulatory Approaches to LLM Governance

## 3  Regulating Bias in LLMs

The current landscape of AI-generated content is subject to several regulatory frameworks, although none of them have been devised with the possibility in mind that LLMs may affect citizens' fundamental worldviews and social perspectives, especially their political opinions and voting decisions. In this section, we explain how these frameworks can nonetheless specifically address communication bias in LLMs, and what their constraints and limitations are.

### 3.1 The AI Act

The AI Act contains a set of obligations that apply to AI systems and GPAI models, among other things, and which explicitly encompass LLMs (Art. 51(1) point (a); Annex XIII AI Act). In essence, AI models are the technological foundation, while AI systems are the applications built on top of them (e.g., ChatGPT, which adds a user interface, moderation tools, and specific workflows). The AI Act defines an AI system as a machine-based system designed to operate with varying levels of autonomy and which is capable of generating outputs such as predictions, content, or recommendations based on input data (Art. 3(1) AI Act). By contrast, GPAI models are those trained with large amounts of data using self-supervision and capable of performing a wide range of tasks (Art. 3(63) AI Act). Providers of AI systems must meet requirements tailored to the system's use context, such as risk management, ensuring high-quality and representative training data, and transparency for high-risk applications. Providers of GPAI models must comply with specific obligations. They must prepare technical documentation, share information with downstream system providers, and publish a summary of training data. There are additional requirements if the model poses a systemic risk (due to high capability or widespread use), including model evaluation, adversarial testing, systemic risk assessment, and incident reporting. When a provider integrates its model into a system and offers it directly, both sets of obligations—those for models and for systems—apply (Recitals 85, 97 AI Act) [2, 12].

Several of these obligations are relevant in addressing bias in LLMs. For high-risk systems in particular, high-quality training data must be used, and developers must implement data governance measures to ensure that training datasets are representative and do not perpetuate biases. This includes regular audits of datasets used for training LLMs. The AI Act considers biases to be relevant if they are likely to affect the health and safety of persons, negatively impact fundamental rights, or lead to discrimination prohibited under EU law, especially where data outputs influence inputs for future operations (Art. 10(2) AI Act). Moreover, the AI Act requires providers to establish a continuous risk management system that identifies, analyzes, and mitigates foreseeable risks to health, safety, and fundamental rights, including risks from bias in system outputs (Art. 9 AI Act). This process necessitates regular monitoring, testing (including under real-world conditions), and targeted measures to reduce residual risks that directly pertain to communication bias in LLMs. The AI Act complements this obligation by requiring providers to ensure their systems achieve and maintain appropriate levels of accuracy and robustness, explicitly requiring measures to manage feedback loops that could introduce or amplify bias over time, and to benchmark, document, and transparently declare accuracy metrics (Art. 15 AI Act). Technical solutions must also address vulnerabilities that could be exploited to manipulate outputs, which includes bias-related exposure. In addition, the AI Act requires that human oversight mechanisms are in place for high-risk AI systems, ensuring that outputs can be reviewed and corrected if they exhibit biased or harmful features (Art. 14 AI Act). Such oversight is crucial for maintaining accountability in decision-making processes influenced by LLMs [35]. Similarly, some developers must actively prevent LLMs from producing biased or discriminatory outputs. This includes conducting impact assessments to evaluate how the models' outputs affect different social groups and to ensure that harmful or biased content generated by LLMs can be reported and addressed effectively, and corrective actions can be taken if biases are identified (Art. 27 AI Act).

### 3.2 The DSA

Several of the DSA's provisions may have implications for LLMs, although their application is more nuanced. The DSA regulates a broad range of digital services, focusing on intermediary and hosting service providers as well as online platforms. This includes search engines and recommender or advertising systems, which may indirectly encompass services utilizing LLMs. The core components of the DSA mandate that platforms implement mechanisms to address illegal content, ensure transparency in their content moderation practices, enable users to challenge content moderation decisions, and hold platforms accountable for illegal or harmful content disseminated through their services. If LLMs are used for content generation, the platforms utilizing LLMs must comply with content moderation and transparency obligations.

The DSA imposes stricter obligations on VLOPs or VLOSEs that reach more than 45 million active EU service recipients on average each month and are designated as such under the DSA (Art. 33(1) DSA). While many LLMs do not meet this threshold on their own, if they are integrated into VLOPs or VLOSEs (for example, as part of a chatbot feature on a social media website), they may be captured by the DSA. VLOPs and VLOSEs must, among other things, identify, analyze, and mitigate systemic risks associated with their services, particularly those related to illegal content, fundamental rights, freedom of expression and information—including the freedom and pluralism of the media—and any actual or foreseeable negative effects on civic discourse and electoral processes through the design of their recommender and/or other relevant algorithmic systems, their content moderation mechanisms, data-related practices, and so forth.

However, the DSA does not explicitly target LLMs as standalone entities, and its compliance requirements are far from clear-cut. It remains contentious whether GPAI models such as LLMs fall within the DSA's definitions, and such classifications hinge upon, among other things, interpretations of how LLMs store and provide user inputs and outputs [15]. While LLMs that are part of online platforms might be the best fit for the DSA's provisions, most generative AI products will not qualify as VLOPs or VLOSEs as they typically do not disseminate information publicly in a manner consistent with traditional online platforms.

### 3.3 Limitations of Existing Legal Frameworks

We have identified several legal obligations that potentially ensure that LLMs do not perpetuate bias in digital information ecosystems; however, hardly any of them can conclusively deal with communication bias.

The AI Act contains several legal requirements—such as regular auditing, transparency, and responsible development practices—to help ensure that LLMs do not perpetuate bias in digital information ecosystems, but these measures have notable limitations in directly addressing communication bias. While the AI Act requires pre-market auditing (Arts. 9, 10, 43 AI Act) to identify and mitigate bias, including the use of diverse and representative training data and technical documentation, its obligations focus mainly on the provider's side and are most robust before LLMs are deployed. While transparency measures (Arts. 13, 53 AI Act) mandate clear documentation of training data, model architecture, and risks, to enhance user awareness, these measures do not directly target or reduce the propensity of LLMs to generate biased outputs. In addition, for GPAI models presenting systemic risk—typically limited to a handful of advanced LLM providers—the AI Act imposes further obligations, including mandatory model evaluation, adversarial testing, systemic risk assessment, serious incident reporting, and adequate cybersecurity protections (Arts. 51, 55 AI Act). However, these requirements largely retain the AI Act's pre-deployment focus, leaving post-deployment bias mitigation chiefly in the hands of providers rather than enabling ongoing, independent, or user-driven measures. Post-market, the AI Act requires the implementation of monitoring measures (Art. 72 AI Act), albeit such measures are largely provider-driven and reactive, relying on incident reporting and periodic audits rather than real-time, user-guided bias correction. Deployers have certain operational duties, such as monitoring system usage and reporting incidents (Arts. 26, 61, 73 AI Act), yet their role in the ongoing mitigation of biases remains relatively limited and is largely passive rather than proactive. Additional constraints—such as strict data protection rules on processing sensitive data for bias detection (Art. 9 GDPR, Art. 10(5) AI Act)—further restrict the scope of alleviating bias. By and large, the AI Act's strongest mechanisms for bias correction converge in the pre-deployment phase, with more reflexive and less specific post-deployment safeguards [16, 38].

While in our view these strategies are essential for fostering trust in digital information ecosystems, we emphasize the need to further explore how a more balanced environment for content generation and dissemination can be promoted, especially once LLMs have been placed on the market. We address this issue in the next section.

## 4   WHERE DO WE GO FROM HERE?

For the most part, the AI Act and the DSA lack provisions that specifically address communication bias in LLMs; where the AI Act and the DSA nonetheless apply, any mitigation of communication bias will typically be a byproduct rather than an intended objective of this legislation. This is partly because communication bias in LLMs can be subtle, multifaceted, and nuanced, rendering it difficult to target and evaluate such bias objectively [3, 14, 26, 27]. In addition, striking the right balance between fostering innovation and implementing effective regulation depends on an ongoing evaluation of the impacts of LLMs on society. In the following paragraphs, we describe some conceptual challenges specific to 1) value chain regulation; and 2) content moderation. We then 3) propose putting a greater emphasis on market mechanisms aimed at aligning technological design with societal values as a complement to existing regulations; and 4) suggest several concrete measures to implement our proposal.

### 4.1   Conceptual Challenges Arising from Value Chain Regulation

Value chain regulation of LLMs seeks to sensibly address the risks associated with AI-driven technologies, but such regulation also presents several challenges that can impact the models' development, deployment, and overall effectiveness. For instance, value chain regulation can impose overly strict requirements that may stifle innovation and discourage developers from experimenting with new approaches or technologies, slowing down advancements in LLM capabilities and applications. Owing to the rapid advancement of AI technologies, regulators often struggle to keep pace with new developments, and existing frameworks may quickly become outdated and/or fail to address pertinent issues [2]. Indeed, compliance with new regulations can be overwhelming, especially for smaller companies or startups. Navigating regulatory obligations may require significant resources, diverting attention from core development activities and potentially leading to reduced competitiveness in the market. A one-size-fits-all attitude can lead to inappropriate obligations that fail to adequately address the harms associated with different use cases [29]. Value chain regulation also leads to the possibility that large companies can influence regulatory outcomes to their advantage, leading to provisions that favor established players over smaller actors [33]. This can stifle competition within the sector and pose challenges when monitoring compliance across different platforms or applications is required. As LLMs transition from development to widespread deployment, it becomes increasingly critical to consider not only how the law can mitigate biases before LLMs are released to the market but also how a balanced and pluralistic information environment can be ensured once they are in active use.

### 4.2 Conceptual Challenges Arising from Content Moderation

Content moderation is often considered essential in addressing communication bias, but content moderation mechanisms also raise concerns around freedom of expression [17]. This is partly due to the law's ambiguous language and the potential for such rules to be misused by authorities. For instance, the AI Act defines systemic risk as a threat unique to the high-impact capabilities of GPAI models, which can significantly affect the EU market due to their extensive reach or actual or anticipated effects on public health, safety, and fundamental rights, including upon society at large (Art. 3(65) AI Act). The DSA, for its part, identifies a broad spectrum of risks, such as the spread of illegal content and negative effects on fundamental rights, civic discourse, public safety, and issues related to gender-based violence and the well-being of minors (Art. 34(1) DSA). There are good reasons to protect these interests; however, it is relatively easy to envisage how application of such rules might come to restrict rather than promote freedom of expression. For instance, authorities have employed safety, public security, and public health interests as reasons to justify speech restrictions [36]. In other instances, concerns around public safety and national security have been raised to defend shutdowns of the Internet [39]. Content moderation rules are often framed in general and sweeping terms and are subject to different interpretations. In addition, enforcement of the provisions is often overseen by a political body—in the EU, the EC via the AI Office—which has been viewed as an additional risk to freedom of expression. Similar issues pertain to algorithmic content moderation mechanisms. Despite their scalability relative to human moderation, the use of algorithms presents distinctive challenges due to the algorithms' ability to produce conflicting predictions, their arbitrariness in output generation, disparate treatment of different datasets, and overall lack of consistency and predictability [14]. In addition to addressing overt harms, we therefore argue that the law should support ongoing efforts that promote a diversity of viewpoints and equitable access to information, enabling a more balanced digital ecosystem for content generation and dissemination as LLMs become progressively entrenched in public discourse.

### 4.3 Competition and Ongoing Technology Design Governance as Complements to Existing Regulations

Value chain regulation and content moderation focus on pre- and post-market decision-making relating to harmful content. This may, however, act as a distraction, diverting attention from a closer examination of what the underlying incentives of LLMs are and how they introduce and generate communication bias in the first place. In particular, content moderation focuses on the outputs of LLMs, removing pieces of illegal content and banning individuals or groups that disseminate unwanted information. An assumption underlying content moderation is that social media produces toxic user-generated subject matter because malicious actors exploit technology products to inflict harm on others, while the technology is considered neutral, simply mirroring the undesirable characteristics of users. However, when it comes to addressing subtle communication bias, content moderation rules reveal several shortcomings. Historically, technology companies operated under an approach that emphasized consumer sovereignty, where the focus was on providing users with tools to express their viewpoints freely [22]. In the interim, there has been a notable shift toward an approach in which AI systems implement guardrails to prevent the dissemination of harmful or contentious output. This raises issues about what constitutes harmful content and who decides which output ought to be suppressed. To illustrate, some AI systems have been found to refuse to generate specific information that opposes certain social justice positions while readily producing other material that supports them [26]. This presents challenges for existing legal frameworks. Governments might exert pressure on technology companies, via content-based regulations, (not) to promote particular political agendas, complicating the background of free speech and regulatory oversight. Furthermore, the concentration of media power makes it feasible for authorities to target platforms based on their influence, raising ethical and legal questions about censorship and the role of private companies in public discourse. Content moderation may become polarizing due to the inevitability that particular opinions will be quashed (some, perhaps, to a greater extent than others) because such opinions are liable to violate the policies of certain platforms. This has led, in part, to the recent move by Meta to end its fact-checking programs and to replace professional content moderators with user-generated community notes [34]. Content moderation mechanisms thus face an inherent trade-off—the apprehension is that strict enforcement will affect more posts made by innocent citizens, which will accidentally be taken down, and vice versa. This exposes a fundamental flaw in the expectation that profit-driven platforms can effectively self-regulate, especially as changing societal and political pressures mount.

As AI systems become gatekeepers of information, LLMs may reinforce existing biases and promote particular perspectives. This may lead to a less pluralistic political environment where dissenting opinions are marginalized, making it more difficult for users to encounter diverse viewpoints. The potential for AI outputs to influence public opinion is particularly concerning in closely contested political landscapes. We therefore propose to focus on competition and ongoing technology design governance as a complement to value chain and content-based regulation. Contrary to conventional wisdom, it is well recognized in scientific literature that the properties of technology products help shape the behavior of human actors and that technology products are not blank slates that simply expose users' behavior [4, 19, 22, 32]. Instead, the design of technology products often drives the behavior of human actors, similar to the way in which computer code can function as law. Technology design determines, at least to some extent, what users see, the information they divulge, the level of privacy they enjoy, and how they interact with others. Technology design can persuade and subtly nudge individual users toward specific actions and often seeks to maximize engagement rather than encourage behavior that benefits democratic integrity and public trust in particular sources of information [39]. Our proposal therefore involves putting a greater emphasis on regulation that helps to effectively readdress the incentives that drive the deployment of AI technologies and ensures, from the outset, that technology platforms proceed in the best interests of consumers [22]. Government regulation that aims to place guardrails on AI companies may suppress specific outputs that are considered harmful or immoral, forcing those companies to assume editorial roles, which may ultimately limit consumer sovereignty. This raises issues about who decides what constitutes harmful content, whether such content reflects a biased worldview that can stifle legitimate discourse, and whether the established guardrails may themselves introduce bias into AI outputs. As the recent move by Meta to end its fact-checking program demonstrates, such regulation may make AI companies even more susceptible to government pressure, because officials might target firms to promote particular agendas [34]. What is more, because of the unique characteristics of generative AI technologies, the imposition of content-related mandates may impede rather than facilitate their ability to produce meaningful and relevant outputs, and the complexity involved in moderating content might contradict, rather than promote, the desired level of discretion and contextual understanding of diverse topics that the development of public opinion typically demands [36].

Moreover, if a few large AI companies progressively wield a disproportionate influence over political information, such concentration can lead to a homogenization of perspectives in which alternative points of view risk becoming marginalized or excluded. Beyond value chain regulation and content moderation—which allow some intervention on the basis of truth or falsity and require an obligation of expression to ensure that communication is not misleading—we highlight the promotion of competition in the AI market. Low barriers to entry can dilute the influence of dominant players and facilitate a more diverse range of outputs, mitigating the risks associated with concentrated power and communication bias in a few LLMs [18, 40]. This is not to neglect the risk that increasing the number of outputs may introduce malicious and bad-faith actors, or to claim that the promotion of competition achieves virtually the same as debiasing. Accentuating these risks, however, conflates two distinct issues: defining the boundaries of public discourse in ways that give proper scope to some indispensable values (i.e., reducing bias at the source through value chain regulation and targeting illicit and harmful outcomes by way of content moderation); and fostering a more diverse public sphere in which a plurality of viewpoints can be articulated and debated. The promotion of competition does not seek to multiply sources of bias but aims to ensure that public discourse includes the full range of communicative processes deemed necessary for the formation of public opinion [31]. In particular, where the removal of content is inherently contentious because such removal risks being perceived as censorship and presumes that LLMs should act as editorial gatekeepers, competition must play an essential part in addressing communication bias. And competition can further alleviate such bias when multiple models achieve similar overall accuracy but lean in different directions in specific instances, or have a disproportionate effect on some users or groups compared to others [5, 6, 30].

However, promoting competition among multiple AI systems should not simply result in AI companies' pursuit to capture market shares; rather, competition should aim toward enabling the fulfillment of a broad spectrum of communicative needs and values in the public sphere. Achieving this objective involves addressing moral hazard—a situation where technology companies face minimal accountability for the risks they impose on users [25]. The moral hazard in communication bias stems from misaligned incentives between corporations seeking to maximize attention and engagement and users expecting to be able to rely on digital ecosystems that provide them with the most accurate and unbiased information [8]. This means that the promotion of competition must simultaneously incorporate a significant extent of user control in terms of self-governance—that is, where users can have a say in the design of AI systems in terms of data collection/cleaning, model architecture/training and output evaluation/refinement throughout their entire lifecycle—and help shape public opinion where LLMs can potentially influence communicative action [23]. Where users are allowed to determine, at least in part, the parameters that define the content they wish to generate or access through AI technologies, they are authorized to better express their expectations and viewpoints freely, without undue interference. Affording users a meaningful path of influence through continuous self-governance in synergy with competition will then offer a more robust response to communication bias in LLMs than can be achieved by relying solely on value chain regulation or content moderation. In an environment in which platforms' business models are driven, almost exclusively, by incentives to maximize engagement and attention, user preferences in relation to diversity, inclusivity, and integrity are routinely sidelined. Establishing a robust infrastructure that ensures meaningful user participation in the ongoing operation of LLMs may then help realign market incentives to permit AI companies to better serve the broader public interest [21, 25].

Some of the EU Digital Market Act's (DMA) provisions serve that broader purpose, affording users a greater number of options in the market and compelling firms to internalize the harms resulting from data-driven technologies [24]. On the one hand, the DMA establishes a set of clearly defined criteria to identify gatekeepers—large undertakings providing core platform services (CPSs) such as online intermediation services, online search engines, online social networking services, video-sharing platform services, messaging services, operating systems, web browsers, virtual assistants, cloud computing services, and online advertising services (Arts. 2 and 3 DMA). While LLMs are not explicitly referenced as CPSs, they could potentially be included in the DMA's list of CPSs. In addition, LLMs may serve as foundational technologies that enhance or drive such services, such as when they are used in search engines for generating responses or summarizing information. Furthermore, companies that provide LLMs may rely on CPSs to deploy their models more effectively. For instance, an LLM might be integrated into a messaging service or a social media platform to facilitate user interactions. If a company providing LLMs also operates a CPS, it would be subject to the obligations contained in the DMA [7, 28]. On the other hand, the DMA includes several obligations that can contribute to bias mitigation in LLMs. For instance, under the Act, gatekeepers are prohibited from favoring their services or products over those of competitors to unfairly disadvantage competing services. By prohibiting the practice of self-preferencing, the DMA encourages a level playing field for new technologies—an equal chance to succeed by requiring all actors to operate by the same set of rules—and a more diverse range of AI services to be made available to users, reducing the dominance of a single model that may exhibit particular biases [20]. The DMA also includes obligations related to how data is combined across different services provided by gatekeepers. These obligations require transparency and user consent when combining data from various sources. By regulating how data is combined and ensuring that users have control over it, the DMA mitigates bias arising from specific datasets. If LLMs are trained on more diverse and representative data, inherent biases in outputs will decrease. Moreover, the DMA prohibits gatekeepers from using, in competition with business users, non-public data generated or provided by those business users in the context of their use of the relevant CPSs, including data generated or provided by the customers of those business users. When applied to LLMs, this means that gatekeepers are prohibited from utilizing non-public data to train their AI models and to ensure that all providers have equal access to specific data sources. This further mitigates bias arising from training models on proprietary or exclusive datasets that fail to represent the broader population. Moreover, by requiring gatekeepers to rely on publicly available data, stakeholders can better scrutinize the sources of information that influence model behavior and can better identify and address potential biases.

Preventing gatekeepers from leveraging non-public data will also help level the playing field for LLM developers, to the extent that smaller companies and new entrants can compete effectively against larger firms that might otherwise dominate due to exclusive access to valuable data. Lastly, the DMA mandates that under certain circumstances, gatekeepers must provide third parties with access to essential data and services. This obligation may be crucial in fostering competition and allowing smaller actors and new entrants to access the data they require in order to develop new LLMs.

### 4.4 Recommendations

Although existing regulations such as the AI Act and DSA focus on mitigating risks and harmful outputs, they often overlook the equally important need to promote diversity (competition) and representativity (self-governance) in the design of digital technologies [15]. We highlight three sets of measures that can better ensure diversity and representativity in ongoing technology design to address communication bias in LLMs.

In the first instance, existing legal provisions should be interpreted and applied with the possibility of communication bias in mind. To illustrate, regular auditing should include assessing how LLMs align with particular perspectives. Developing benchmark datasets, such as the GermanPartiesQA [3], can help evaluate the alignment of LLMs with political party positions and detect biases in their responses. Understanding how LLMs exhibit sycophancy—where the models tailor responses to align with user opinions—can help mitigate bias. By conducting prompt experiments that analyze how LLM outputs change based on user prompts, developers can better understand and address the models' propensities. In providing documentation about the training data, model architecture, and decision-making processes of LLMs, users should be educated about the potential biases inherent in LLMs and how these biases may affect generated content. Ensuring that training data includes a wide range of viewpoints, demographics, and cultural contexts can further mitigate the risk of reinforcing stereotypes. Existing post-deployment obligations establish a regulatory basis for enhancing the responsiveness of AI systems through built-in human oversight tools, intervention protocols, and active monitoring of AI outputs. In addition, the rapid adoption and market impact of LLM-based applications justify the inclusion of LLMs within the regulatory scope of the DMA. Not only could existing provisions be applied to LLMs (especially when operated by designated gatekeepers), but LLMs could also be covered by amending the DMA to explicitly include LLMs as distinct CPSs. This would mandate compliance with DMA obligations and enable the EC to designate LLM providers as gatekeepers if they meet the DMA's quantitative or qualitative thresholds. Finally, current post-deployment mechanisms for mitigating communication bias in LLMs under the AI Act are limited by narrow complaint procedures that do not empower users to meaningfully influence product design or the generation of AI outputs. To institutionalize more effective complaint and investigation procedures, oversight mechanisms should move beyond provider self-assessment and internal monitoring, and instead grant authorities broader investigative and remedial powers. Investigations should involve establishing external audits, enabling authorities to proactively examine systemic risks, and requiring providers to implement product design changes on behalf of users, where necessary. Drawing on the DSA's framework, vetted external researchers and auditors could be given access to LLMs to assess bias and the effectiveness of proposed mitigation strategies, with findings informing specific interventions. Additionally, authorities should be empowered to act based on user complaints, with the ability to put binding enforcement measures in place so that users can articulate and put into effect their apprehensions. This would shift oversight from a pure pre-market approach to an inclusive, participatory market-centered model that better protects public discourse by addressing all aspects of communication bias in LLMs.

## 5 CONCLUSION

Value chain regulation and content moderation exhibit a tendency to favor snapshot assessments at specific points in time and to focus on the front- or back-end of the algorithmic design process. Competition, together with an increased level of self-governance, highlights an additional requirement for content generation and dissemination once LLMs have been placed on the market. This is particularly pertinent in terms of bias mitigation since competition and self-governance establish conditions that are aimed at facilitating an active search for those LLMs that are the most accurate and least partial, shifting responsiveness upstream to designers and overseers and compelling them to engage in a continuous evaluation and improvement of AI systems. Since the potential of LLMs comes with multiple issues and ambiguities, a number of different regulatory approaches are required to enable citizens to have evenhanded access to the public sphere and effectively partake in the creation of public opinion. We hope our overview will offer a distinct perspective and highlight important means by which the development of AI-based technologies can be propelled in useful directions.

**ACKNOWLEDGMENTS**

We would like to thank the German Research Foundation (DFG), SPP 2378 (project ReNO), 2023-2027, the Weizenbaum Institute for the Networked Society for a generous fellowship grant, the HKU Faculty Development Fund, the HKU Seed Fund for Basic Research, project code: 2401102107, and the HKU URC Funding Scheme.

**REFERENCES**

[1] AI phone: Deutsche Telekom wants to free smartphones from apps, February 2024. Available at https://www.telekom.com/en/media/media-information/archive/deutsche-telekom-frees-smartphones-from-apps-1060272.

[2] M. Almada and N. Petit. The EU AI Act: Between the rock of product safety and the hard place of fundamental rights, *Common Market Law Review*, 62:85, 2025.

[3] J. Batzner, V. Stocker, S. Schmid, G. Kasneci. GermanPartiesQA: Benchmarking commercial large language models and AI companions for political alignment and sycophancy, *8th AAAI/ACM Conference on AI, Ethics, and Society (AIES)*, Madrid, Spain, October 2025.

[4] L. Bennett Moses and M. Zalnieriute. Law and technology in the dimension of time. In S. Ranchordás and Y. Roznai, editors, *Time, Law, and Change: An Interdisciplinary Study*. Hart, 2020.

[5] E. Black, J.L. Koepke, P.T. Kim, S. Barocas, M. Hsu. Less discriminatory algorithms. *Georgetown Law Journal*, 113:53, 2024.

[6] E. Black, M. Raghavan, S. Barocas. Model multiplicity: Opportunities, concerns, and solutions. In *Proceedings of the 2022 ACM conference on fairness, accountability, and transparency*, 850, 2022.

[7] F. Bostoen and J. Krämer. AI agents and ecosystems contestability. Technical report, CERRE, November 2024.

[8] S. Ennis. Consumer expectations and fair contracting for digital products. CPI antitrust chronicle, July 2022.

[9] Z. Epstein and A. Hertzmann. Art and science of generative AI. Understanding shifts in creative work will help guide AI's impact on the media ecosystem. *Science*, 380, June 16 2023.

[10] European Union Agency for Cybersecurity. ENISA threat landscape 2023, October 2023.

[11] Federal Bureau of Investigation. Internet crime report, Internet crime complaint center, 2023.

[12] D. Fernández-Llorca, E. Gómez, I. Sánchez, G. Mazzini. An interdisciplinary account of the terminological choices by EU policymakers ahead of the final agreement on the AI Act: AI system, general purpose AI system, foundation model, and generative AI, *Artificial Intelligence and Law*, 2024.

[13] U. Gasser and V. Mayer-Schoenberger. *Guardrails: Guiding Human Decisions in the Age of AI*. Princeton University Press, 2024.

[14] J.F. Gomez, C. Machado, L. Monteiro Paes, F. Calmon. Algorithmic arbitrariness in content moderation. February 26 2024.

[15] R. Gorwa. *The Politics of Platform Regulation. How Governments Shape Online Content Moderation*. Oxford University Press, 2024.

[16] Ş.İ. Göksal, M.C. Solarte Vasquez, A. Chochia. The EU AI Act's alignment within the European Union's regulatory framework on artificial intelligence, *International and Comparative Law Review*, 24: 25, 2024.

[17] P. Hacker, A. Engel, M. Maurer. Regulating ChatGPT and other large generative AI models. May 12 2023.

[18] A. Jones and W.E. Kovacic. Antitrust's implementation blind side: Challenges to major expansion of U.S. competition policy. *Antitrust Bulletin*, 65:227, 2020.

[19] B.J. Koops and R. Leenes. Privacy regulation cannot be hardcoded. A critical comment on the 'privacy by design' provision in data-protection law, *International Review of Law, Computers & Technology*, 28: 159, 2013.

[20] A. Kuenzler. Advancing quality competition in big data markets. *Journal of Competition Law and Economics*, 15:500, 2019.

[21] A. Kuenzler. Competition law as a catalyst for collective market governance: Gauging the discursive benefits of intensified administrative action. *Yearbook of European Law*, 41:252, 2022.

[22] A. Kuenzler. *Restoring Consumer Sovereignty. How Markets Manipulate Us and What the Law Can Do About It*. Oxford University Press, 2017.

[23] A. Kuenzler. The shadow of digital antitrust: The law's purported deference to leading actors' product design choices. *Concurrences Review*, 4:1, 2022.

[24] A. Kuenzler. Third-generation competition law. *Journal of Antitrust Enforcement*, 11:133, 2023.

[25] A. Kuenzler. What competition law can do for data privacy (and vice versa). *Computer Law & Security Review*, 47:1, 2022.

[26] T. Ma. LLM echo chamber: Personalized and automated disinformation, September 2024.

[27] T. Margoni, J. Quintais, S. Schwemer. Algorithmic propagation: How the data-platform regulatory framework may increase bias in content moderation. In C. Sganga and T. Synodinou, editors, *Flexibilities in Copyright Law*. Routledge, 2025.

[28] A. Martinez. Generative AI in check: Gatekeeper power and policy under the DMA, 2024. Available at https://papers.ssrn.com/sol3/papers.cfm?abstract_id=5025742.

[29] M.G. Martinez Alles. *Torts and Retribution. The Case for Punitive Damages*. Cambridge University Press, 2025.

[30] S. Passi and S. Barocas. Problem formulation and fairness. In *Proceedings of the 2019 conference on fairness, accountability, and transparency*, 39, 2019.

[31] R. Post. Participatory democracy and free speech. *Virginia Law Review*, 97:477, 2011.

[32] K. Prifti. The theory of 'regulation by design': Towards a pragmatist reconstruction. *Technology and Regulation*, 152, 2024.

[33] Q.B. Schäfer. AI, IP, and competition policy: Adjusting policy levers to a new GPT. In R. Abbott and T. Schrepel, editors, *Artificial Intelligence and Competition Policy*. Concurrences, 2024.

[34] M. Isaac and T. Schleifer. Meta to end fact-checking program in shift ahead of Trump term. *The New York Times*, January 7 2025. Section Technology.

[35] N.A. Smuha and K. Yeung. The European Union's AI Act: Beyond motherhood and apple pie? In N.A. Smuha, editor, *The Cambridge Handbook of the Law, Ethics and Policy of Artificial Intelligence*, Cambridge, 2025.

[36] R. Tushnet. Content moderation in an age of extremes. *Journal of Law, Technology and the Internet*, 10:1, 2019.

[37] UK Department for Science, Innovation and Technology. A pro-innovation approach to AI regulation: Government response. Command paper: CP 1019, February 2024. Available at: https://www.gov.uk/government/consultations/ai-regulation-a-pro-innovation-approach-policy-proposals/outcome/a-pro-innovation-approach-to-ai-regulation-government-response.

[38] S. Wachter. Limitations and loopholes in the EU AI Act and AI liability directives: what this means for the European Union, the United States, and beyond. *Yale Journal of Law & Technology*, 26:671, 2024.

[39] R. Wilson and M. Land. Hate speech on social media: Content moderation in context. *Connecticut Law Review*, 52:1029, 2021.

[40] A.G. Yasar, A. Chong, E. Dong, T.K. Gilbert, S. Hladikova, R. Maio, C. Mougan, X. Shen, S. Singh, A-A. Stoica, S. Thais, M. Zilka. AI and the EU Digital

Markets Act: Addressing the risks of bigness in generative AI, July 2023.